\documentclass[aps,pre,twocolumn,superscriptaddress]{revtex4}
\usepackage{amsmath, amssymb}
\usepackage{graphicx}
\usepackage{latexsym}
\usepackage{esint}

\begin{document}

\title{Onset of negative interspike interval correlations in adapting neurons}

\author{Eugenio Urdapilleta}
\email[]{urdapile@ib.cnea.gov.ar} \affiliation{Divisi\'on de
F\'isica Estad\'istica e Interdisciplinaria \& Instituto Balseiro,
Centro At\'omico Bariloche, Avenida E. Bustillo Km 9.500, S. C. de
Bariloche (8400), R\'io Negro, Argentina}

\begin{abstract}
Negative serial correlations in single spike trains are an
effective method to reduce the variability of spike counts. One of
the factors contributing to the development of negative
correlations between successive interspike intervals is the
presence of adaptation currents. In this work, based on a hidden
Markov model and a proper statistical description of conditional
responses, we obtain analytically these correlations in an
adequate dynamical neuron model resembling adaptation. We derive
the serial correlation coefficients for arbitrary lags, under a
small adaptation scenario. In this case, the behavior of
correlations is universal and depends on the first-order
statistical description of an exponentially driven
time-inhomogeneous stochastic process.
\end{abstract}

\maketitle

\section{Introduction}
\indent Spike-frequency adaptation (SFA) is one of the main
adaptation mechanisms in neural systems \cite{kandel, wark2007}.
As its name implies, the effect defining SFA is the observation of
a scaling in the input-output relationship between the injected
current (or a stimulus property) and the firing rate of a spiking
neuron, from an initial to a stationary mapping
\cite{ermentrout1998, benda2003, higgs2006, prescott2008}. Several
mechanisms can contribute to SFA (for example, depressing synapses
\cite{chung2002}); however, the most prominent mechanism
accounting for SFA is the presence of (probably simultaneous)
spike-related currents, which produce a negative feedback to the
neuron in time scales ranging from tens to thousands of
milliseconds \cite{benda2003, madison1984, helmchen1996, sah1996,
hille}. Even when the full impact of these currents on neural
coding is not completely understood, it is known that they
contribute to the processing of static as well as temporal
signals. For the processing of temporal signals, it is worth
pointing out the high-pass filtering characteristics due to SFA
\cite{benda2003, benda2005, benda2010} and its related sensitivity
to input fluctuations \cite{higgs2006}, the forward masking effect
\cite{wang1998, liu2001}, the selectivity to complex stimuli
\cite{peron2009a, peron2009b}, and the enhanced reliability of
temporal coding \cite{prescott2008}.\\
\indent For the case of static signals, given the absence of a
temporal structure in the input, a neural system makes use of a
rate code to represent them. A rate code is defined as the number
of spikes in a certain temporal window and it is completely
described by the spike count statistics \cite{dayan}.
Spike-related adaptation currents have a twofold impact on this
code. First, they modify the tuning curve between signals and
responses or the input-output relationship mentioned above, which
helps to match dynamic ranges \cite{wark2007, ermentrout1998,
benda2003, higgs2006, prescott2008}. Second, they introduce
negative correlations between successive interspike intervals
(ISIs) in stationary neural spike trains \cite{prescott2008,
benda2010, wang1998, liu2001, avila2011}. While the first effect
reflects the modification of the \textit{mean} spike count, the
second effect implies a strong change in its \textit{variance}
(and higher-order properties). This change arises from the fact
that the presence of negative correlations in a point process
reduces the long-term variability in the counting process that
defines the rate code \cite{cox}. Taken together, both effects
deeply affect the encoding reliability \cite{avila2011,
nawrot2007, nawrot2010, farkhooi2009, farkhooi2011}. The presence
of negative correlations also affects the coding capabilities of
other related schemes; for example, coding of slowly varying
signals through a mechanism called \textit{noise shaping}
\cite{chacron2004, lindner2005, avila2009} (demonstrated for
negative correlations arising from a history-dependent threshold,
but also valid for spike-related adaptation currents), or
adaptation-based independent codes \cite{nesse2010}.\\
\indent Correlated single spike trains constitute a nonrenewal
point process. In neural systems, this kind of process is
relatively ubiquitous \cite{lowen, lowen1992, longtin1997,
ratnam2000, chacron2001} (for reviews, mostly based on negative
correlations, see also \cite{avila2011, farkhooi2009}). In
general, there is a coexistence of processes that evoke opposite
effects on the ISI correlations, and therefore on the counting
statistics in single neurons (adaptation currents and other
regularizing processes such as synaptic depression and negative
feedback versus filtering and input correlations, among others).
Such interesting scenarios have attracted relative attention
within the theoretical neuroscience community, and several studies
focus on these statistics or related properties in different
situations \cite{farkhooi2009, farkhooi2011, lindner2005,
nesse2010, chacron2001, middleton2003, lindner2004a,
schwalger2008a, schwalger2010a}. Related with our present work, we
should note three approaches that have been derived recently: a
population-based scheme for adapting ensembles \cite{muller2007}
(see also \cite{farkhooi2009, farkhooi2011, nesse2010}), a
directed discrete representation for counting events with a
general internal structure \cite{schwalger2010b} (see also
\cite{lindner2007, schwalger2008b}), and a general framework for
nonrenewal processes as a hidden Markov model \cite{vreeswijk}. In
particular, our work can be framed within the
general approach obtained by van Vreeswijk \cite{vreeswijk}.\\
\indent In this work, based on the results we have found in a
previous study about the first-passage-time (FPT) problem in an
exponentially driven Wiener process \cite{urdapilleta}, we derive
how the resulting negative correlations arise in the spike train
of a dynamical, although simple, process resembling the basic
features of a spike-related adaptation current added to a spiking
neuron in the presence of fast additive fluctuations. The
expressions we find are strictly valid in a slight adaptation
regime, where the complete FPT density is not necessary and
perturbation techniques can be applied, so they remain valid for
other dynamical models (with the spike-related adaptation current
considered here, or similar). In this way, the onset of
correlations due to adaptation is general across different models,
provided the additive noise is fast. In the final part of the
work, we use the statistics of the FPT problem and the emerging
correlations due to adaptation to assess how the spike count
variance decreases in comparison to an equivalent neuron without
adaptation, in an asymptotic limit. This reduction in the spike
count variance underlies the improvement in the decoding
performance, and we show how the dependence on the correlations
and on the intrinsic variability reduction due to the
inhomogeneous driving shape the spike count variance reduction for
different spiking frequencies.

\section{Theoretical framework}
\subsection{Basic model}
\indent We consider an integrate-and-fire (IF) neuron, where
subthreshold dynamics of the membrane potential $V$ is governed by

\begin{equation}\label{eq1}
   C_{\text{m}}\frac{dV}{dt} = f(V) + I_{\text{adapt}} + I_{\text{ext}}.
\end{equation}

\indent External, $I_{\text{ext}}$, as well as internal currents,
$f(V)$ and $I_{\text{adapt}}$, drive the membrane potential
whenever $V < V_{\text{thr}}$. The internal current $f(V)$ takes
into account different interspike phenomena, such as leakage or
spike initiation onset. The simplest models are the perfect [$f(V)
= 0$] and the leaky [$f(V) = -g_{L} V$] IF neurons, where only
leakage is considered (the perfect IF model corresponds to no
leakage). The subthreshold dynamics is supplemented by a threshold
condition, which simplifies the highly nonlinear process of a
spike excursion: whenever the potential reaches $V_{\text{thr}}$ a
spike is defined and immediately, the membrane potential is set to
a reset condition $V_{\text{r}}$.\\
\indent Several types of adaptation currents have been
characterized by experiments \cite{madison1984, helmchen1996,
sah1996}. According to previous theoretical studies \cite{liu2001,
benda2003, benda2010}, we model the adaptation current as a
process $x(t)$ that decays during spikes and is incremented when a
spike event occurs. In particular, we consider the adaptation
current as $I_{\text{adapt}}(t) = - g_{\text{a}}~x(t)$, where the
interspike dynamics for the adaptation process is given by

\begin{equation}\label{eq2}
   \frac{dx}{dt} = -\frac{x}{\tau_{\text{a}}},
\end{equation}

\noindent and a fixed increase $\alpha>0$ is evoked at all spike
times $t_{\text{sp}}$: $x(t_{\text{sp}}) \rightarrow
x(t_{\text{sp}}) + \alpha$. This model could be considered as an
idealization of the Ca$^{2+}$-dependent K$^{+}$ current, which is
widely expressed in neurons exhibiting SFA [$x(t)$ would represent
the calcium concentration in a current-based scheme]. Without
mathematical loss, we can set $g_{\text{a}} =
C_{\text{m}}/\tau_{\text{a}}$ and use $\alpha$ to
control the strength of the adaptation current.\\
\indent Since $I_{\text{adapt}}(t) = -
(C_{\text{m}}/\tau_{\text{a}})~x(t)$, from Eq.~(\ref{eq2}) it
follows that, between the ($n$)th and the ($n+1$)th spikes, the
evolution of the adaptation current is given by

\begin{equation}\label{eq3}
   \frac{I_{\text{adapt}}(t)}{C_{\text{m}}} = - \frac{\varepsilon_{\text{n}}}{\tau_{\text{a}}}~\exp[-(t-t_{\text{n}})/
   \tau_{\text{a}}],
\end{equation}

\noindent where $\varepsilon_{\text{n}}$ represents the state of
the adaptation process $x$ immediately after the spike time
$t_{\text{n}}$.\\
\indent Without a random component, the deterministic dynamics of
the membrane potential in the IF model with an adaptation current,
Eqs.~(\ref{eq1}) and (\ref{eq3}), evolves along a prescribed
trajectory and no correlations emerge since each ISI is a replica
of itself (however, even in the deterministic regime,
perturbations propagate in the sequence of ISIs, which induces
correlations \cite{schwalger2010a}). We introduce a stochastic
component in the system through external noise. In particular, we
assume that the external current is given by a constant
deterministic component and an additive Gaussian white noise
representing fast external fluctuations

\begin{equation}\label{eq4}
   \frac{I_{\text{ext}}}{C_{\text{m}}} = \mu + \xi(t),
\end{equation}

\noindent where $\langle \xi(t) \rangle = 0$ and $\langle \xi(t)
\xi(t') \rangle = 2D \delta(t'-t)$. Therefore, the subthreshold
dynamics between the ($n$)th and the ($n+1$)th spikes is
determined by

\begin{equation}\label{eq5}
   \frac{dV}{dt} = g(V) + \mu - \frac{\varepsilon_{\text{n}}}{\tau_{\text{a}}}~\exp[-(t-t_{\text{n}})/\tau_{\text{a}}] +
   \xi(t).
\end{equation}

\begin{figure*}[t]
\begin{center}
\includegraphics[scale=0.525]{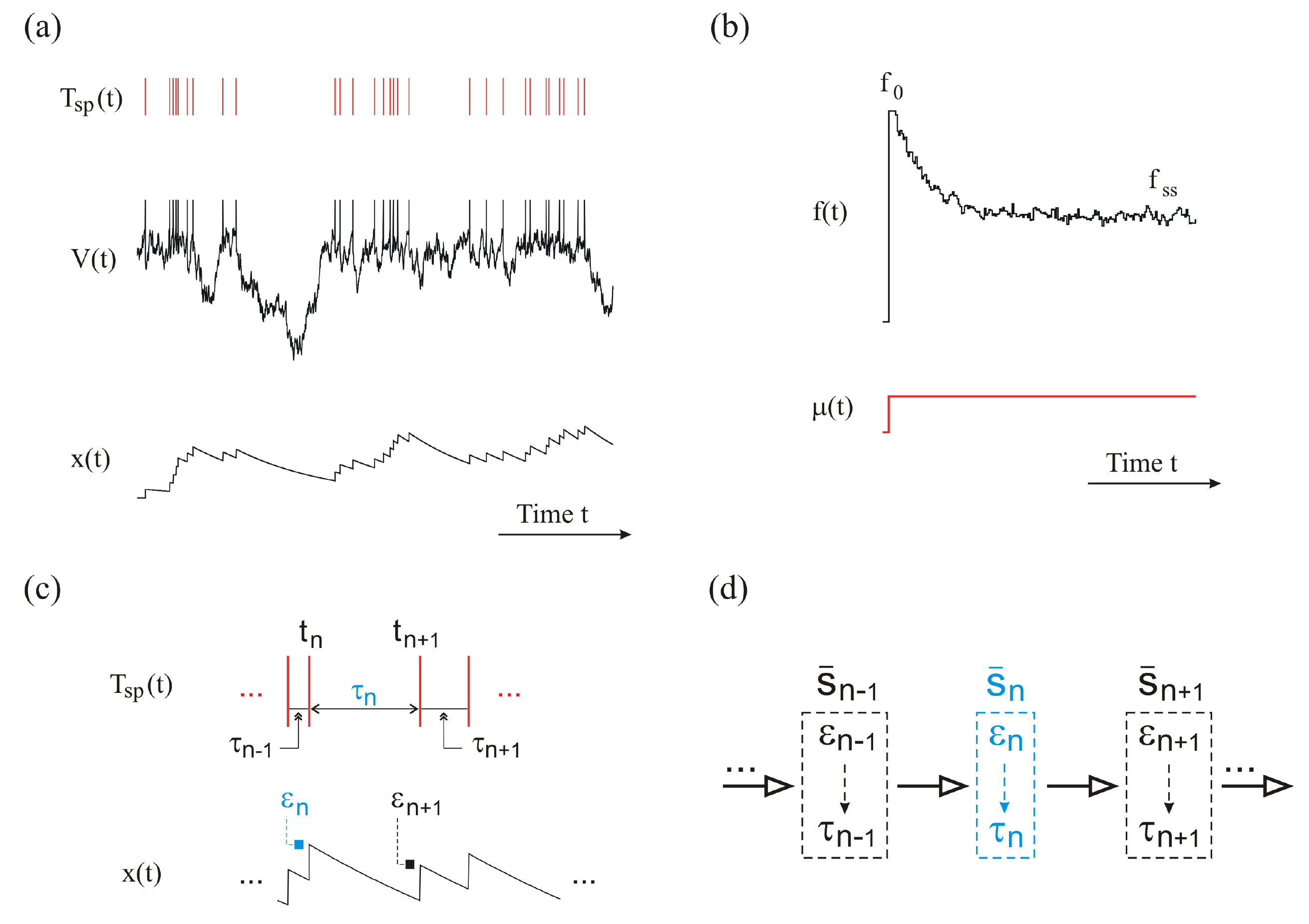}
\caption{\label{fig1} (Color online) Neuron model with adaptation
current. (a) A system realization, $V(t)$ and $x(t)$, and the
resulting spike train $T_{\text{sp}}(t)$. (b) The system develops
spike-frequency adaptation (top), usually characterized by the
temporal evolution of the firing frequency, $f(t)$, stimulated
with step currents (bottom). (c) A sample trace of the adaptation
process $x(t)$ and the basic elements used to describe the
stochastic nature of the system, $\varepsilon$ and $\tau$. (d) The
hidden Markov model, composed by $\varepsilon$ and the observable
$\tau$, used for the analysis.}
\end{center}
\end{figure*}

\indent Different (IF) models will have different $g(V)$
functions. In particular, $g(V) = 0$ for the perfect IF model and
$g(V) = -V/\tau_{\text{m}}$, with $\tau_{\text{m}} =
C_{\text{m}}/g_{L}$, for the leaky IF model. The threshold
condition (at spike time $t_{\text{n}+1}$) resets the membrane
potential to $V_{\text{r}}$ and updates the adaptation state to
$\varepsilon_{\text{n+1}} =
\varepsilon_{\text{n}}~\exp(-\tau_{\text{n}}/\tau_{\text{a}})+\alpha$,
where $\tau_{\text{n}} = t_{\text{n}+1}-t_{\text{n}}$ is the
($n$)th ISI.\\
\indent In Fig.~\ref{fig1}(a) we show a typical realization of the
membrane potential, $V(t)$, and the adaptation process, $x(t)$,
for the system described above. As shown in Fig.~\ref{fig1}(b),
this model exhibits SFA, where a step input (bottom) induces an
initial firing rate $f_{\text{0}}$ which decays to a lower
steady-state firing rate $f_{\text{ss}}$ (top), with some typical
time scale. Spike times $t_{\text{n}}$ defining the spike train,
$T_{\text{sp}}(t) = \sum \delta(t-t_{\text{n}})$ [see top of
Fig.~\ref{fig1}(a)], also establish the sequence of FPT processes,
$\{ \dots, \tau_{\text{n-1}}, \tau_{\text{n}}, \tau_{\text{n+1}},
\dots\} \equiv \{ \tau_{\text{n}} \}$ [Fig.~\ref{fig1}(c)]. The
ongoing (initial) state of the adaptation process,
$\varepsilon_{\text{n}}$, is a history-dependent random variable
turning the sequence $\{ \tau_{\text{n}} \}$ into non-Markovian.

\indent However, the bidimensional state $\overline{s} =
(\varepsilon,\tau)$, defined at spike times, supports a Markovian
process since the state at (spike) time $t_{\text{n}}$ is
completely characterized from the knowledge of the state at
(spike) time $t_{\text{n-1}}$: given $\varepsilon_{\text{n-1}}$
and $\tau_{\text{n-1}}$, $\varepsilon_{\text{n}}$ is defined by
the deterministic relationship $\varepsilon_{\text{n}} =
\varepsilon_{\text{n-1}}~ \exp(-\tau_{\text{n-1}}/
\tau_{\text{a}}) + \alpha$, and $\tau_{\text{n}}$ is given by the
FPT problem of a certain stochastic process with an exponential
time-dependent drift $(-\varepsilon_{\text{n}}/\tau_{\text{a}})
\exp[-(t-t_{\text{n}})/\tau_{\text{a}}]$. In particular, for the
perfect (leaky) IF model, the underlying stochastic process is a
Wiener (Ornstein-Uhlenbeck) diffusion process. In mathematical
terms, the transition probability density is given by

\begin{eqnarray}\label{eq6}
   f(\overline{s}_{\text{n}} |
   \overline{s}_{\text{n-1}},\overline{s}_{\text{n-2}},\dots) \hspace{4.0cm} \nonumber\\
    = f(\varepsilon_{\text{n}},\tau_{\text{n}} | \varepsilon_{\text{n-1}},\tau_{\text{n-1}})
    \hspace{3.65cm}
   \nonumber\\
    = \delta\{\varepsilon_{\text{n}} - [\varepsilon_{\text{n-1}}
   \exp(-\tau_{\text{n-1}}/\tau_{\text{a}}) +
   \alpha]\}~\phi(\tau_{\text{n}}|\varepsilon_{\text{n}}),
\end{eqnarray}
\\
\noindent where $\phi(\tau_{\text{n}}|\varepsilon_{\text{n}})$ is
the FPT density function associated to the ($n$)th ISI, which
depends exclusively on $\varepsilon_{\text{n}}$ and not on
previous outcomes ($t_{\text{n}}$ in the above notation for the
drift just set the initial time). This Markovian process is
represented in Fig.~\ref{fig1}(d) and constitutes a hidden Markov
model. Based on Eq.~(\ref{eq6}), the key elements necessary to
analyze this stochastic system are the statistics of $\varepsilon$
and the time-inhomogeneous FPT density function.

\subsection{Statistics of the (initial) adaptation strength}
For one-dimensional systems as the IF models, the solution to the
FPT problem is given as an expansion in terms of the strength of
the exponential drift \cite{urdapilleta, comment1}

\begin{equation}\label{eq7}
   \phi(\tau|\varepsilon) = \sum_{\text{i}=0}^{\infty}
   \varepsilon^{\text{i}}~\phi_{\text{i}}(\tau).
\end{equation}

\indent Given the transition density, Eq.~(\ref{eq6}), the
equilibrium probability density for $\varepsilon$,
$\rho_{\text{eq}}(\varepsilon)$, should satisfy \cite{vreeswijk}

\begin{equation}\label{eq8}
   \rho_{\text{eq}}(\varepsilon_{\text{n}}) = \int
   f_{\varepsilon}(\varepsilon_{\text{n}}|\varepsilon_{\text{n-1}})~
   \rho_{\text{eq}}(\varepsilon_{\text{n-1}})~d\varepsilon_{\text{n-1}},
\end{equation}

\noindent where the transition density between substates is

\begin{figure}[t]
\begin{center}
\includegraphics[scale=0.55]{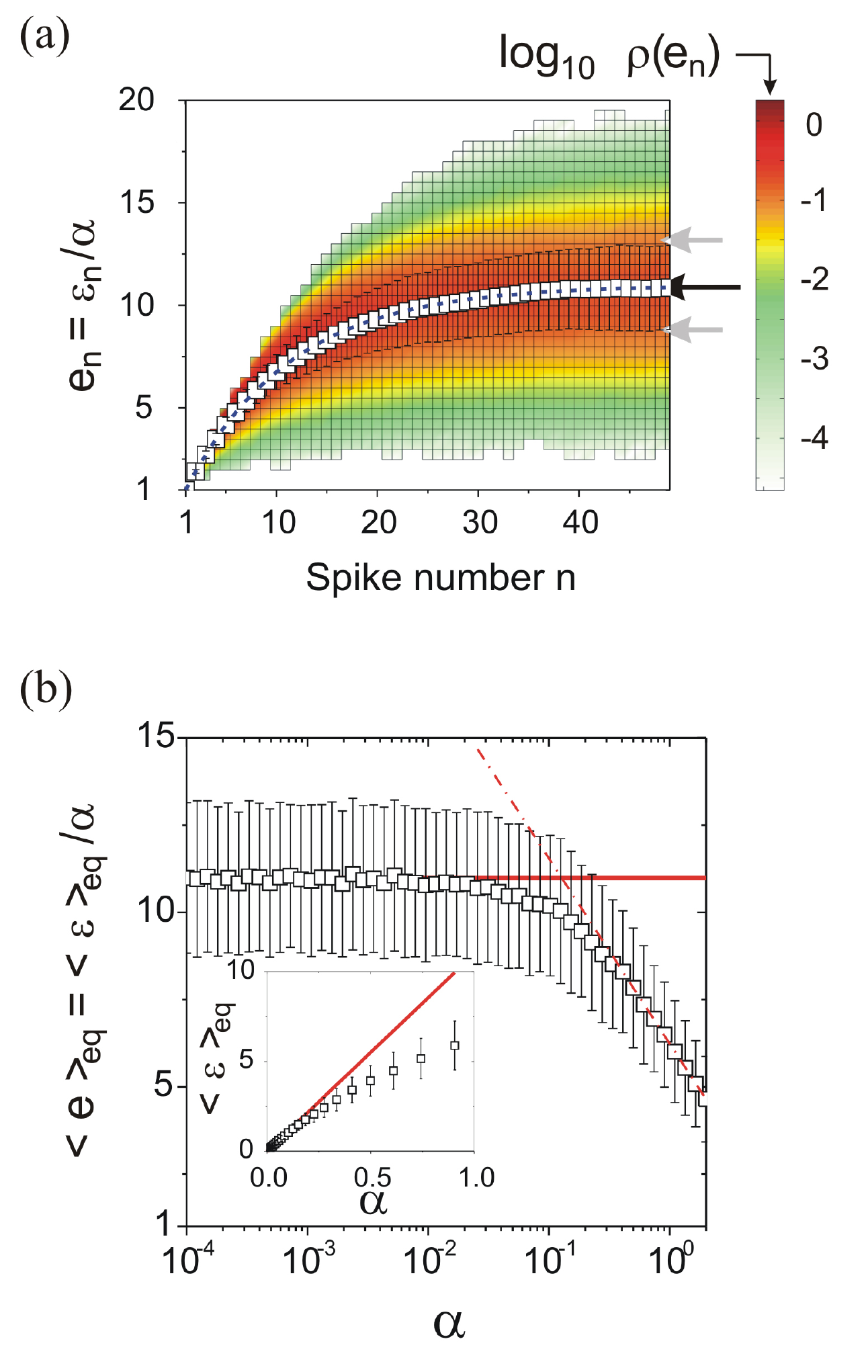}
\caption{\label{fig2} (Color online) Statistics of $\varepsilon$
for the perfect IF model. (a) Density of normalized $\varepsilon$
as a function of the spike number (color bar in logarithmic
scale). Symbols and error bars indicate mean and standard
deviation, respectively. The analytical expression for the mean as
a function of the spike number, Eq.~(\ref{eq16}), is represented
by the dotted (blue) line. The analytical expressions for the
stationary properties (mean/standard deviation), Eqs.~(\ref{eq11})
and (\ref{eq12}), are indicated by arrows at the right
(black/gray). Numerical results were obtained for $\mu = 0.10
~\text{ms}^{-1}$, $D = 0.05 ~\text{ms}^{-1}$,
$V_{\text{thr}}-V_{\text{r}} = 1.0$, $\tau_{\text{a}} = 100.0
~\text{ms}$, and $\alpha = 0.01$ (number of trials,
$N_{\text{trials}} = 10^5$). (b) The normalized stationary mean as
a function of $\alpha$ (semilogarithmic plot). The analytical
result is represented by a continuous (red) line. Nonlinear
effects for large $\alpha$ are fitted by a (dashed-dotted) line,
indicating that linear regime remains valid up to $\alpha \approx
0.1$. Inset: plot of the (not normalized) stationary mean as a
function of $\alpha$. Same parameters as in (a), except that
$N_{\text{trials}} = 10^3$ for each point.}
\end{center}
\end{figure}

\begin{equation}\label{eq9}
f_{\varepsilon} (\varepsilon_{\text{n}} |
\varepsilon_{\text{n}-1}) = \int
f(\varepsilon_{\text{n}},\tau_{\text{n}}
|\varepsilon_{\text{n-1}},\tau_{\text{n-1}})
\phi(\tau_{\text{n-1}}|\varepsilon_{\text{n-1}}) d\tau_{\text{n}}
d\tau_{\text{n-1}}.
\end{equation}

\indent Replacing Eq.~(\ref{eq6}) and the $\varepsilon$-expansion
for $\phi(\tau_{\text{n-1}}|\varepsilon_{\text{n-1}})$,
Eq.~(\ref{eq7}), in Eq.~(\ref{eq9}) we obtain a self-consistent
integral equation, hard to tackle analytically. However, from this
integral equation it is relatively simple to find a relationship
between the moments, which reads

\begin{eqnarray}\label{eq10}
\langle \varepsilon_{\text{n}}^{\text{m}} \rangle & = & \int
\varepsilon_{\text{n}}^{\text{m}}
~\rho_{\text{eq}}(\varepsilon_{\text{n}})~d\varepsilon_{\text{n}}
\nonumber\\
 & = & \alpha^{\text{m}} + \sum_{\text{i}=1}^{\text{m}}
\binom{\text{m}}{\text{i}} ~\alpha^{\text{m}-\text{i}}
~\sum_{\text{j}=0}^{\infty}
\phi_{\text{j}}^{L}(\text{i}/\tau_{\text{a}}) ~\langle
\varepsilon_{\text{n-1}}^{\text{i}+\text{j}}\rangle,
\end{eqnarray}

\noindent where $\binom{a}{b}$ is the binomial coefficient and
$\phi_{\text{j}}^{L}(s)$ is the Laplace transform of the $j$th
term in the expansion for the FPT solution. The previous
relationship relates unconditional moments (subindexes are
irrelevant), and therefore it represents a set of infinite
algebraic equations for the (infinite) moments $\langle
\varepsilon^{\text{m}} \rangle$.\\
\indent In the slight adaptation regime, $\alpha \neq 0$ but
small, it is easy to see that the moments $\langle
\varepsilon^{\text{m}} \rangle \sim
\mathcal{O}(\alpha^{\text{m}})$. In particular, the first two
moments read

\begin{eqnarray}\label{eq11}
   \langle \varepsilon \rangle & = & \frac{\alpha}{[1-\phi_{0}^{L}(1/\tau_{\text{a}})]}, \\
   \label{eq12}
   \langle \varepsilon^{2} \rangle & = & \frac{\alpha^{2} ~
   [1+\phi_{0}^{L}(1/\tau_{\text{a}})]}{ [1-\phi_{0}^{L}(1/\tau_{\text{a}})]~[1-\phi_{0}^{L}(2/\tau_{\text{a}})]}.
\end{eqnarray}

\indent The $\alpha$-normalized properties, $(1/\alpha) \langle
\varepsilon \rangle$ and $(1/\alpha^{2}) \langle (\varepsilon
-\langle \varepsilon \rangle)^2 \rangle$, are shown in
Fig.~\ref{fig2}(a) in black and gray arrows, respectively (right
margin), for the case of the perfect IF neuron model. As expected,
they agree with the results obtained from simulations in the
asymptotic limit (symbols plus error bars: mean plus standard
deviation of $\varepsilon_{\text{n}}$; colored histogram:
logarithm of the density).\\
\indent For small $\alpha$ values, there is an additional
interesting viewpoint to derive $\langle \varepsilon \rangle$,
also valid for any neuron model. Multitrial experiments usually
start from rest, where we assume that there is no adaptation,
$\varepsilon_{0} = 0$. Conditional to this fact, the ($n$)th event
satisfies

\begin{equation}\label{eq13}
   \varepsilon_{\text{n}} = \alpha ~[ 1 + \sum_{\text{i}=1}^{\text{n}-1} \prod_{\text{j}=\text{i}}^{\text{n}-1} \exp(-\tau_{\text{j}}/\tau_{\text{a}}) ].
\end{equation}

\indent The first moment is given by

\begin{equation}\label{eq14}
   \langle \varepsilon_{\text{n}} \rangle = \alpha~[1+\sum_{\text{i}=1}^{\text{n}-1} \langle \prod_{\text{j}=\text{i}}^{\text{n}-1}
   \exp(-\tau_{\text{j}}/\tau_{\text{a}}) \rangle ].
\end{equation}

\indent For a weak adaptation process we consider that $\langle
\epsilon_{\text{n}} \rangle \sim \mathcal{O}(\alpha)$, and then
the conditional probabilities required in the above expression are
well approximated by their zero-order expansion,
$f_{\tau}(\tau_{\text{n}} /
\tau_{\text{n-1}},\tau_{\text{n-2}},\dots) =
\phi_{0}(\tau_{\text{n}})$ for all $n$. In this limit it is easy
to see that

\begin{eqnarray}\label{eq15}
   \langle \varepsilon_{\text{n}} \rangle & = &
   \alpha~[1+\sum_{\text{i=1}}^{\text{n}-1}
   \prod_{\text{j}=\text{i}}^{\text{n}-1} \langle
   \exp(-\tau_{\text{j}}/\tau_{\text{a}})\rangle_{\phi_{0}}] \nonumber\\
   & = & \alpha~\sum_{\text{i}=0}^{\text{n}-1}
   [\phi_{0}^{L}(1/\tau_{\text{a}})]^{\text{i}},
\end{eqnarray}

\noindent where $\langle \cdot \rangle_{\phi_{\text{i}}}$
indicates the average with respect to the function
$\phi_{\text{i}}(\tau)$. This geometric series is readily
obtained:

\begin{equation}\label{eq16}
   \langle \varepsilon_{\text{n}} \rangle = \alpha ~\frac{
1-[\phi_{0}^{L}(1/\tau_{\text{a}})]^{\text{n}} }{
[1-\phi_{0}^{L}(1/\tau_{\text{a}})]},
\end{equation}

\noindent and converges asymptotically to
$\alpha/[1-\phi_{0}^{L}(1/\tau_{\text{a}})]$, whenever
$\phi_{0}^{L}(1/\tau_{\text{a}}) < 1$, in concordance with the
previous analysis. The exponential growth of the normalized
adaptation strength $(1/\alpha) \langle \varepsilon_{\text{n}}
\rangle$, predicted by Eq.~(\ref{eq16}), is shown in
Fig.~\ref{fig2}(a) as a function of the spike number $n$ (blue
dotted line), together with the results obtained from numerical
simulations (symbols). The agreement is remarkable for this case
(simulation results were obtained with $\alpha = 0.01$).\\
\indent In order to determine the range of $\alpha$ where the
approximation remains valid, we run several simulations and
calculate the asymptotic (equilibrium) $\langle \varepsilon
\rangle$ for different $\alpha$ values. In Fig.~\ref{fig2}(b) we
show the normalized asymptotic mean value as a function of
$\alpha$ (inset: not normalized mean value). The average exhibits
the linear behavior indicated by the preceding results up to
$\alpha \approx 0.1$, which represents an adaptation of about
$10\%$ [$(f_{0} - f_{\text{ss}})/f_{0}$, see Fig.~\ref{fig1}(c)].

\section{Results}

\subsection{Onset of correlations}

\begin{figure*}[ht!]
\begin{center}
\includegraphics[scale=0.525]{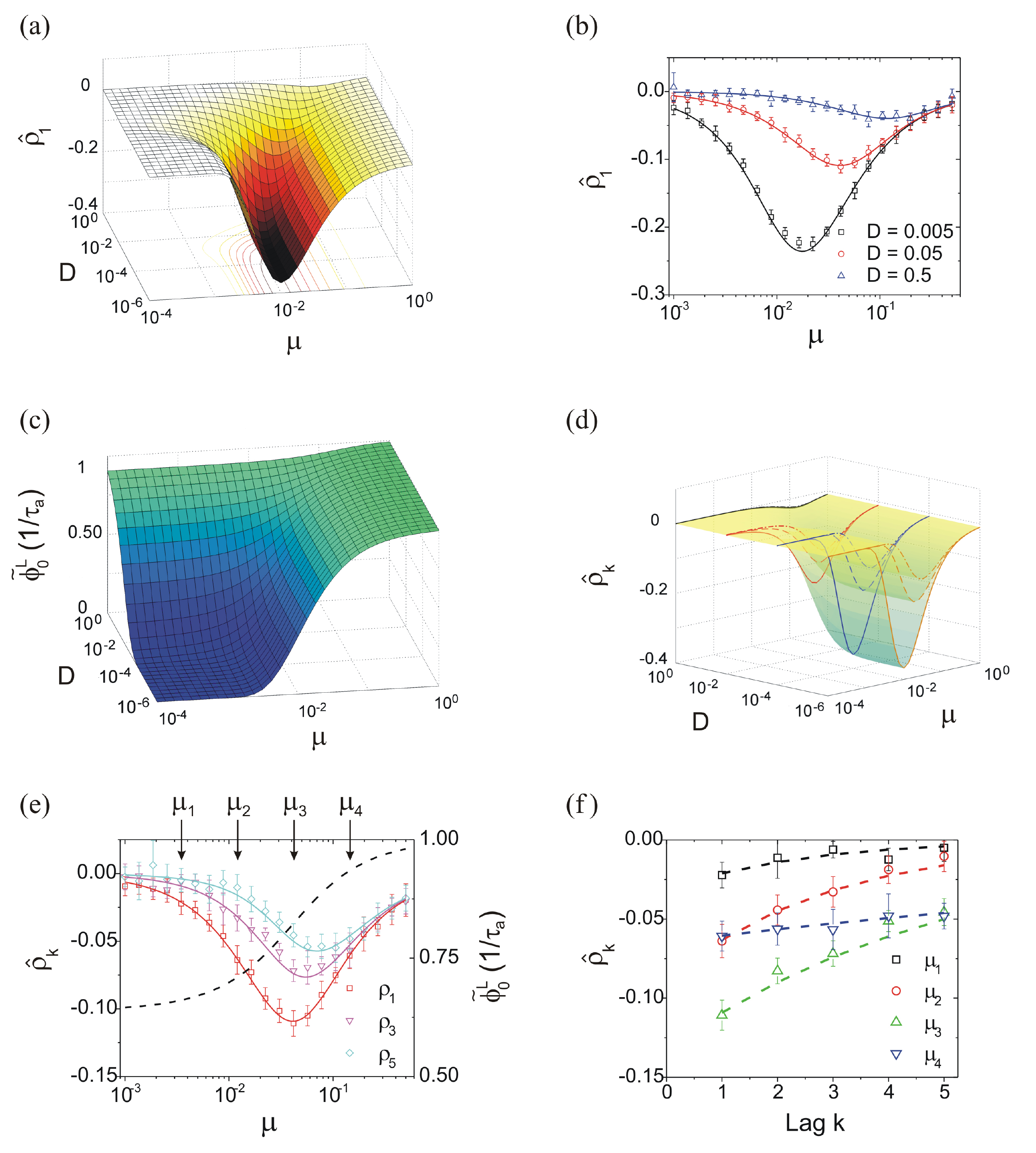}
\caption{\label{fig3} (Color online) Normalized serial correlation
coefficients (SCC), $\hat{\rho}_{k} = \rho_{k}/\alpha$, for the
perfect IF model. (a) Theoretical expression for the normalized
first SCC as a function of driving parameters, $\mu$ and $D$
($\text{ms}^{-1}$) [see Eq.~(\ref{eq23})]. (b) Numerical results
(symbols) for the normalized first SCC as a function of $\mu$
($\text{ms}^{-1}$) and different $D$ values ($\text{ms}^{-1}$).
Corresponding theoretical expressions are indicated by continuous
lines. (c) Factor $\phi_{0}^{L}(1/\tau_{\text{a}})$, which defines
the relationship between successive SCC values, as a function of
the driving parameters. (d) Theoretical prediction of normalized
SCC at higher lags as a function of driving parameters [see
Eq.~(\ref{eq26})]. Continuous, dashed, and dashed-dotted (colored)
lines represent $\hat{\rho}_{1}$, $\hat{\rho}_{2}$, and
$\hat{\rho}_{3}$, respectively, at different noise intensities
(black, red, blue, and orange: $D = 10^{0}$, $10^{-2}$, $10^{-4}$,
and $10^{-6} ~\text{ms}^{-1}$, respectively). Lines are embedded
in their respective continuous (semitransparent) surfaces. (e)
Numerical results for the normalized SCC at higher lags.
$\hat{\rho}_{1}$, $\hat{\rho}_{3}$, and $\hat{\rho}_5$ as a
function of $\mu$ ($\text{ms}^{-1}$, $D = 0.05~\text{ms}^{-1}$).
Continuous lines represent the corresponding theoretical
expressions. Dashed black line indicates
$\phi_{0}^{L}(1/\tau_{\text{a}})$ (right scale). (f) Numerical
results for the normalized SCC as a function of the lag for the
four cases indicated with arrows in (e). Dashed lines indicate
theoretical expressions. Ratios between successive SCCs are given
by the value of $\phi_{0}^{L}(1/\tau_{\text{a}})$ at each point
[dashed line in (e) at the different values of $\mu$]. Numerical
results were obtained from $N_{\text{ISI}} = 10^{6}$ consecutive
ISIs, error bars were estimated from $N_{\text{repet}} = 10$
repetitions. Remaining parameters: $V_{\text{thr}}-V_{\text{r}} =
1.0$, $\tau_{\text{a}} = 100.0 ~\text{ms}$, and $\alpha = 0.1$.}
\end{center}
\end{figure*}

\indent Correlations in nonrenewal point processes are usually
characterized by the serial correlation coefficient (SCC), which
is defined by

\begin{equation}\label{eq17}
   \rho_{\text{k}} = \frac{ \langle \tau_{\text{n}}
   \tau_{\text{n}+\text{k}} \rangle - \langle \tau_{\text{n}} \rangle
   \langle \tau_{\text{n}+\text{k}} \rangle }{ \sqrt{\langle
   \Delta\tau_{\text{n}}^2 \rangle \langle
   \Delta\tau_{\text{n}+\text{k}}^2\rangle} },
\end{equation}

\noindent where $k$ indicates the lag between successive ISIs,
brackets indicate ensemble average, and $\Delta \tau_{\text{i}}^2
= (\tau_{\text{i}}-\langle \tau_{\text{i}} \rangle)^2$ is the
variance. Once the system reaches the stationary conditions (the
firing frequency is adapted), the SCC simplifies to

\begin{equation}\label{eq18}
   \rho_{\text{k}} = \frac{ \langle \tau_{\text{n}}
   \tau_{\text{n}+\text{k}} \rangle - \langle \tau \rangle^2 }{
   \langle \Delta \tau^2 \rangle}.
\end{equation}

\indent For the hidden Markov model defined by Eq.~(\ref{eq6}),
$\langle \tau_{\text{n}} \tau_{\text{n}+\text{k}} \rangle$ reads

\begin{eqnarray}\label{eq19}
   \langle \tau_{\text{n}} \tau_{\text{n}+\text{k}} \rangle =
   \int_{\mathcal{D}s_{\text{n}}} \int_{\mathcal{D} s_{\text{n}+\text{k}}} \tau_{\text{n}} \tau_{\text{n}+\text{k}}~
   f(\varepsilon_{\text{n}+\text{k}},\tau_{\text{n}+\text{k}}|\varepsilon_{\text{n}},\tau_{\text{n}}) \nonumber\\
   \times ~
   \phi(\tau_{\text{n}}|\varepsilon_{\text{n}})~\rho_{\text{eq}}(\varepsilon_{\text{n}})~ds_{\text{n}}~ds_{\text{n}+\text{k}},
\end{eqnarray}

\noindent where $ds_{\text{i}}$ represents
$d\varepsilon_{\text{i}} d\tau_{\text{i}}$ and
$\mathcal{D}s_{\text{i}}$ its integration domain.\\
\indent At lag $\text{k}=1$, the transition probability density
between states is given directly by Eq.~(\ref{eq6}). The
\textit{onset} of correlations is characterized by the smallest
order in $\alpha$ which produce finite SCC values. This order
coincides with the small adaptation regime, and therefore, the
linear $\varepsilon$-expansion suffices for relevant expressions
in Eq.~(\ref{eq19}), $\langle \tau_{\text{n}} \tau_{\text{n}+1}
\rangle \sim \mathcal{O}(\alpha)$. Consequently, we obtain

\begin{eqnarray}\label{eq20}
   \langle \tau_{\text{n}} \tau_{\text{n}+1} \rangle = \langle \tau
   \rangle_{\phi_{0}}^2 \hspace{5.25cm} \nonumber\\
   + ~\alpha ~ \langle \tau
   \rangle_{\phi_{1}} \left[\frac{\langle \varepsilon
   \rangle}{\alpha}\left(\langle \tau \rangle_{\phi_{0}} -
   \frac{d\phi_{0}^{L}(s)}{ds}\rfloor_{1/\tau_{\text{a}}}\right) +
   \langle \tau \rangle_{\phi_{0}}\right],
\end{eqnarray}

\noindent where $\langle \cdot \rangle_{\phi_{\text{i}}}$
indicates the average with respect to the function
$\phi_{\text{i}}(\tau)$. In order to keep the linear order in the
normalization required for the SCC, we have $\langle \tau
\rangle^2 \sim \mathcal{O}(\alpha)$ and $\langle \Delta \tau^2
\rangle \sim \mathcal{O}(0)$:

\begin{eqnarray}\label{eq21}
   \langle \tau \rangle^2 &=& \langle \tau \rangle_{\phi_{0}}^2 + 2
   \langle \varepsilon \rangle \langle \tau \rangle_{\phi_{0}}
   \langle \tau \rangle_{\phi_{1}},\\
   \label{eq22}
   \langle \Delta \tau^2 \rangle &=& \langle
   \Delta\tau^{2}\rangle_{\phi_{0}}.
\end{eqnarray}

\indent In Eqs.~(\ref{eq20}) and (\ref{eq21}), the evaluation of
$\langle \varepsilon \rangle$ should be performed in the linear
regime. Taking into account the results obtained in the previous
section, $\langle \varepsilon \rangle / \alpha =
1/[1-\phi_{0}^{L}(1/\tau_{\text{a}})]$, the first SCC reads

\begin{eqnarray}\label{eq23}
   \rho_{1} = - \alpha~\frac{\langle \tau \rangle_{\phi_{1}}}
   {[1-\phi_{0}^{L}(1/\tau_{\text{a}})] ~\langle \Delta\tau^2
   \rangle_{\phi_{0}}}~\hspace{2cm}\nonumber\\
   \times~\left[ \phi_{0}^{L}(1/\tau_{\text{a}})
   \langle \tau \rangle_{\phi_{0}} + \frac{d\phi_{0}^{L}(s)}{ds}
   \rfloor_{1/\tau_{\text{a}}} \right].
\end{eqnarray}

\indent To evaluate the SCC at higher lags, $\text{k}> 1$, we need
to obtain the transition probability density between the states
$n$ and $n+k$ which, based on the hidden Markov model, reads

\begin{widetext}
\begin{eqnarray}
\label{eq24}
   \begin{split}
   f(\varepsilon_{\text{n}+\text{k}}, \tau_{\text{n}+\text{k}}|\varepsilon_{\text{n}},\tau_{\text{n}})
   &=& \dotsint_{\mathcal{D}s^{(\text{k}-1})} ds^{(\text{k}-1)} ~ f(\varepsilon_{\text{n}+\text{k}}, \tau_{\text{n}+\text{k}},
   \dots,\varepsilon_{\text{n}+1},\tau_{\text{n}+1}|\varepsilon_{\text{n}},\tau_{\text{n}})\\
   &=& \dotsint_{\mathcal{D}s^{(\text{k}-1})} ds^{(\text{k}-1)} \prod_{\text{i}=1}^{\text{k}} f(\varepsilon_{\text{n}+\text{i}},
   \tau_{\text{n}+\text{i}}|\varepsilon_{\text{n}+\text{i}-1},\tau_{\text{n}+\text{i}-1}),
   \end{split}
\end{eqnarray}
\end{widetext}

\noindent where we have simplified the notation to
$ds^{(\text{k}-1)} = ds_{\text{n}+1} \dots
ds_{\text{n}+\text{k}-1}$ and $\mathcal{D}s^{(\text{k}-1)}$ to the
corresponding integration domain. Each of the factors under the
product symbol has the structure given by Eq.~(\ref{eq6}). After
some calculus [it is convenient to leave unevaluated all integrals
in $\tau_{\text{i}}$ in the transition density from
$\overline{s}_{\text{n}}$ to $\overline{s}_{\text{n}+\text{k}}$,
for posterior evaluation in Eq.~(\ref{eq19})], the average between
successive ISIs is given, up to first order in $\alpha$ (and/or
$\langle \varepsilon \rangle$), by

\begin{eqnarray}\label{eq25}
   \langle \tau_{\text{n}} \tau_{\text{n}+\text{k}} \rangle =
   \langle \tau \rangle_{\phi_{0}}^2 ~+~\alpha~\langle \tau
   \rangle_{\phi_{1}}\hspace{3.5cm}\nonumber\\
   \times \Bigg[ \Big( 1+\frac{\langle \varepsilon \rangle}{\alpha} \Big) \langle \tau
   \rangle_{\phi_{0}} + \langle \tau \rangle_{\phi_{0}}
   \sum_{\text{i}=1}^{\text{k}-1} [\phi_{0}^{L}(1/\tau_{\text{a}})]^{\text{k}-\text{i}} \nonumber\\
   - \frac{\langle \varepsilon \rangle}{\alpha}
   [\phi_{0}^{L}(1/\tau_{\text{a}})]^{\text{k}-1}
   \frac{d\phi_{0}^{L}(s)}{ds}\rfloor_{1/\tau_{\text{a}}}\Bigg].
\end{eqnarray}

\noindent Replacing the stationary value for $\langle \varepsilon
\rangle$, and after normalization, it is relatively simple to
prove that SCC values at superior lags are given by

\begin{equation}\label{eq26}
   \rho_{\text{k}} =
   [\phi_{0}^{L}(1/\tau_{\text{a}})]^{\text{k}-1}~\rho_{1}~,
   \hspace{1.0cm} \text{k}>1.
\end{equation}

\indent For the perfect IF model, the lowest orders in the
expansion corresponding to the solution of the FPT problem,
Eq.~(\ref{eq7}), are given \cite{urdapilleta}, and therefore, the
expressions for $\rho_{\text{k}}$ can be explicitly evaluated.\\
\indent In Fig.~\ref{fig3}(a) we show the normalized first SCC,
$\hat{\rho}_{1} = \rho_{1}/\alpha$, as a function of the
parameters governing the dynamics of the perfect IF neuron model,
$\mu$ and $D$. In Fig.~\ref{fig3}(b) we show the theoretical
expression for the normalized SCC $\hat{\rho}_{1}$ as a function
of the driving force $\mu$, for different noise intensities, and
compare it with numerical results. The agreement between both
curves is remarkable, since the limit of small adaptation is
satisfied (numerical results were obtained with $\alpha = 0.1$ and
then normalized).

\begin{figure}[t]
\begin{center}
\includegraphics[scale=0.55]{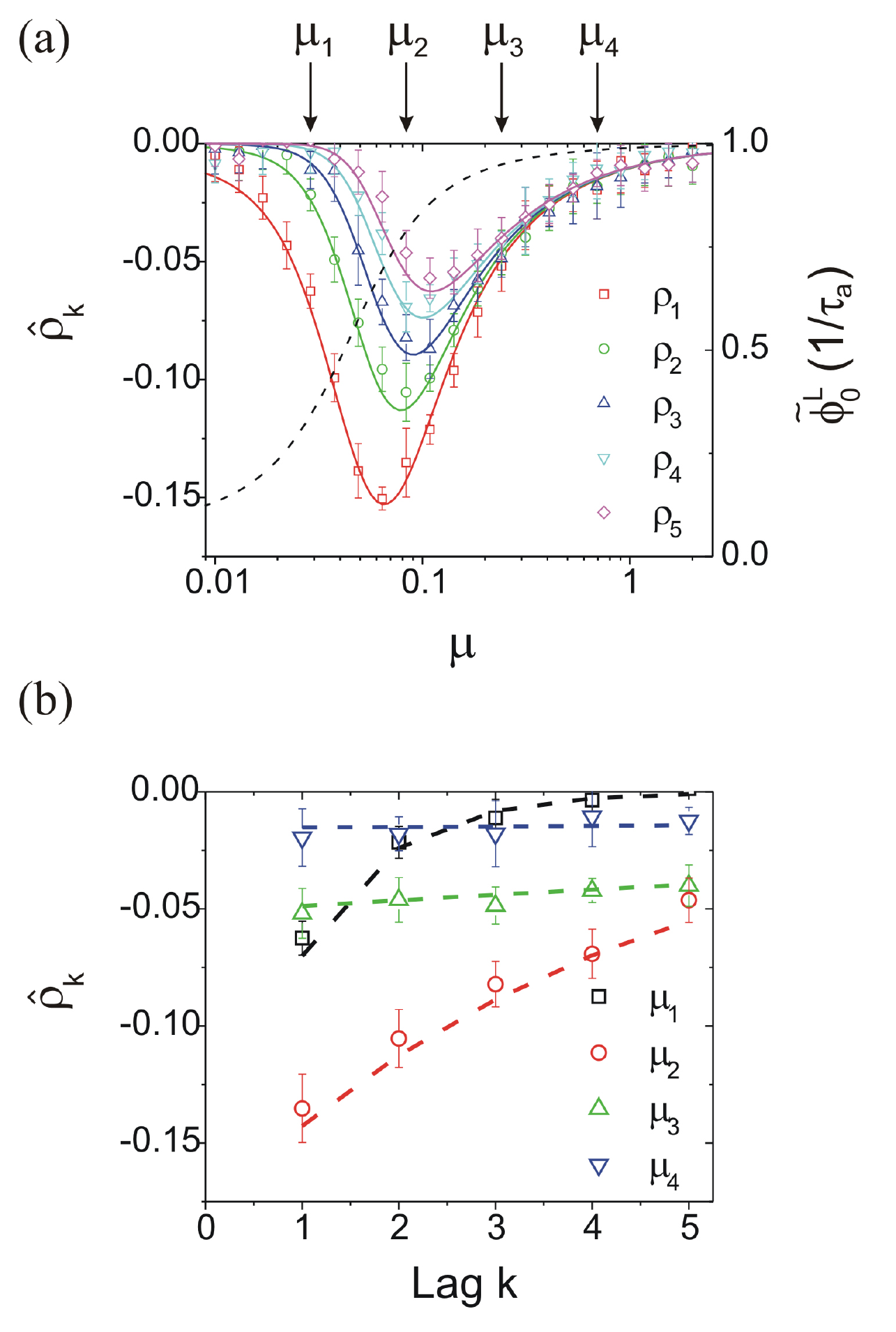}
\caption{\label{fig4} (Color online) Normalized serial correlation
coefficients (SCCs), $\hat{\rho}_{\text{k}} =
\rho_{\text{k}}/\alpha$, for the leaky IF model [$g(V) = -
V/\tau_{\text{m}}$]. (a) and (b) analogous to Figs.~\ref{fig3}(e)
and ~\ref{fig3}(f). Parameters: $\tau_{\text{m}} =
10.0~\text{ms}$, $D = 0.01~\text{ms}^{-1}$,
$V_{\text{thr}}-V_{\text{r}} = 1.0$, and $\tau_{\text{a}} =
100.0~\text{ms}$.}
\end{center}
\end{figure}

\indent As expressed by Eq.~(\ref{eq26}), the SCC at higher lags,
$\rho_{\text{k}}$ for $k>1$, have a geometric structure:
$\rho_{\text{k}} = \phi_{0}^{L}(1/\tau_{\text{a}}) ~
\rho_{\text{k}-1}$. Since $0 < \phi_{0}^{L}(1/\tau_{\text{a}}) <
1$ [see Fig.~\ref{fig3}(c)], SCC at higher lags are scaled
versions of the first SCC. In Fig.~\ref{fig3}(d) we show the first
three normalized SCC, $\hat{\rho}_{\text{k}} =
\rho_{\text{k}}/\alpha$, as a function of the parameters $\mu$ and
$D$. As stated by the geometric relationship, the exact scaling
between them is given according to the precise location in the
parameters space [equivalent point in Fig.~\ref{fig3}(c)]. The
scaling can be better appreciated in Fig.~\ref{fig3}(e), which
simply represents a section of Fig.~\ref{fig3}(d) along a
particular $D$. In this case, theoretical (solid lines) as well as
numerical results (symbols) for $\hat{\rho_{1}}$,
$\hat{\rho_{3}}$, and $\hat{\rho_{5}}$ are represented as a
function of the driving force $\mu$ ($\hat{\rho}_{2}$ and
$\hat{\rho}_{4}$ are omitted for the sake of clarity). The dashed
line shows the respective section of Fig.~\ref{fig3}(c) (scale in
the right margin), which governs the scaling between consecutive
SCCs. Actually, the geometric structure represents an exponential
decay of the SCC as a function of the lag. This can be observed in
Fig.~\ref{fig3}(f) for different $\mu$ values selected in
Fig.~\ref{fig3}(e). The exponential decay is given from the
scaling factor $\phi_{0}^{L}(1/\tau_{\text{a}})$, obtained from
the intersection of the dashed line in Fig.~\ref{fig3}(e) and the
particular value of $\mu$ considered. For example, $\mu_{2}$ and
$\mu_{4}$ were selected so $\hat{\rho}_{1}$ were approximately the
same for both cases. However, the scaling factor
$\phi_{0}^{L}(1/\tau_{\text{a}})$ is higher for $\mu_{4}$ than for
$\mu_{2}$, so the decay is correspondingly slower.\\
\indent The onset of correlations, as characterized by the SCC,
has a general structure, Eqs.~(\ref{eq23}) and (\ref{eq26}), that
relies critically on two factors: the Laplace transform of the
unperturbed distribution, $\phi_{0}^{L}(s)$, and the linear
correction to the mean introduced by the exponential temporal
drift, $\langle \tau \rangle_{\phi_{1}}$. As shown in
\cite{urdapilleta}, for small intensities the main effect of the
exponential drift on the FPT statistics of a perfect IF model is
to change the mean consistently to what we have found in this
work. For other IF models, the complete FPT problem with an
exponential temporal drift can be addressed with a similar
procedure \cite{comment1}. However, even when appealing to set out
the formalism, the complete statistics is not required for
computing the SCC and so we can proceed with simpler approaches.
For example, to compute the linear correction to the mean FPT due
to the exponential temporal drift in generic IF models, we can use
the results obtained by Lindner using a perturbation scheme
\cite{lindner2004b, lindner2005b}. To illustrate the generality of
the results we have found, in Fig.~\ref{fig4} we show the
normalized SCC, $\hat{\rho}_{\text{k}}$, predicted by
Eqs.~(\ref{eq23}) and (\ref{eq26}), using the analytical results,
$\phi_{0}^{L}(s)$ and $\langle \tau \rangle_{\phi_{1}}$, obtained
by Lindner for the leaky IF neuron model \cite{lindner2004b}, and
compare them to numerical simulations. As
expected, theoretical and numerical results agree.\\

\subsection{Loss of linearity}

\begin{figure}[t]
\begin{center}
\includegraphics[scale=0.375]{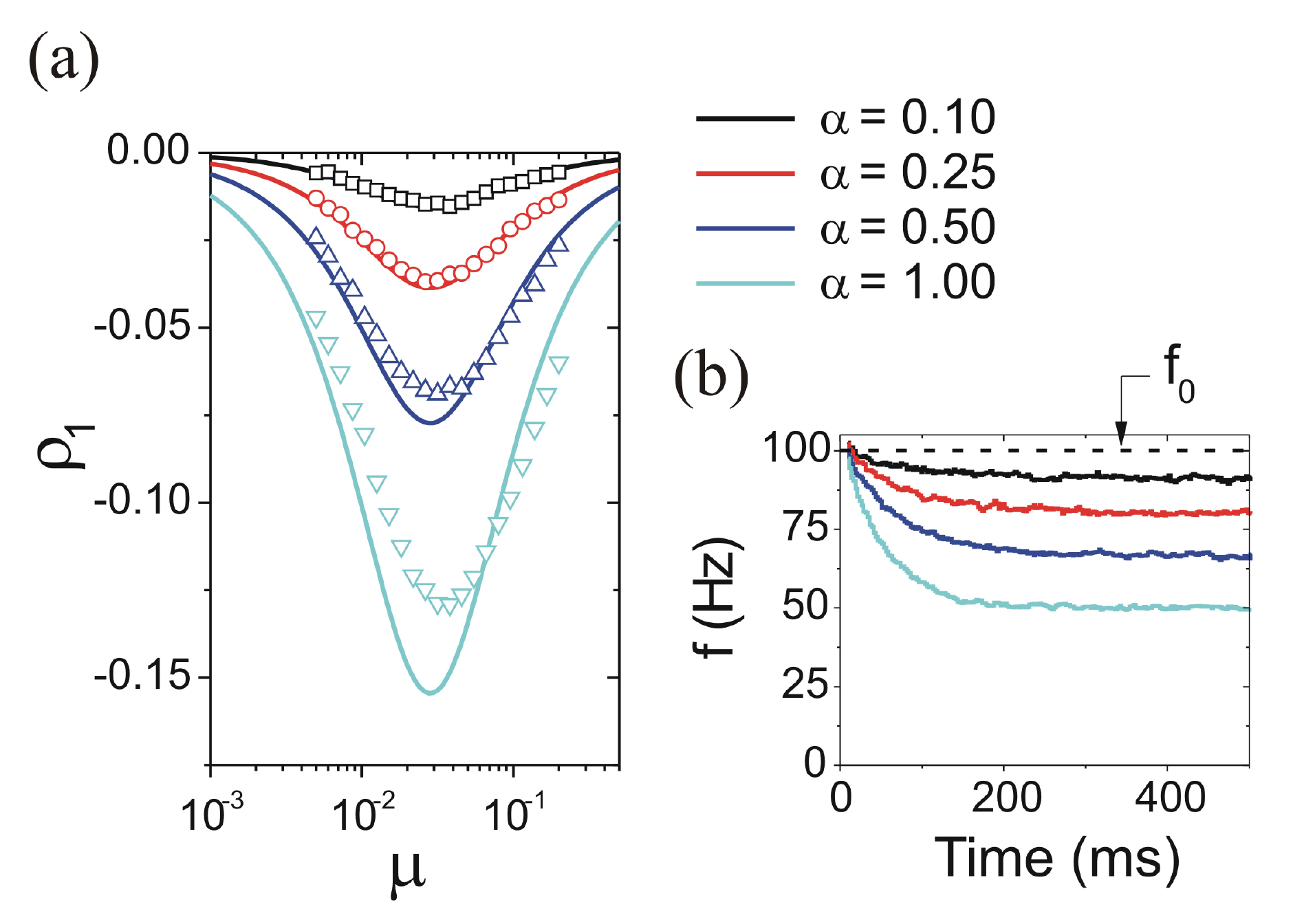}
\caption{\label{fig5} (Color online) Loss of linearity. (a) First
SCC, $\rho_{1}$, as a function of the driving parameter $\mu$ in a
perfect IF neuron model, for different adaptation strengths,
$\alpha$. In all cases, $D = 0.02 ~\text{ms}^{-1}$,
$V_{\text{thr}}-V_{\text{r}} = 1.0$, and $\tau_{\text{a}} = 100.0
~\text{ms}$. (b) As the value of $\alpha$ increases, the
stationary firing rate decreases [asymptotic value of $f(t)$ at
the right margin] and correspondingly, the adaptation effect is
more prominent. In all cases, the driving drift is $\mu =
0.10~\text{ms}^{-1}$ (which corresponds to a firing rate of $100~$
Hz in the absence of adaptation, $\alpha = 0$), applied as a step
function at time $t = 0$ [see Fig.~\ref{fig1}(b)]. Remaining
parameters as in (a).}
\end{center}
\end{figure}

\indent In the previous section we have shown that the onset of
correlations has a specific structure and scales linearly with the
adaptation strength, Eqs.~(\ref{eq23}) and (\ref{eq26}). This
scaling enables us to consider normalized SCC,
$\hat{\rho}_{\text{k}}$, as shown in the preceding figures.
However, for a large enough value of $\alpha$, higher-order
effects become important and these equations are no longer
applicable. In particular, given that the expression for
$\rho_{1}$, Eq.~(\ref{eq23}), does not saturate we expect that
higher-order effects oppose the linear growth. In
Fig.~\ref{fig5}(a) we show the (not normalized) first SCC,
$\rho_{1}$, for different adaptation strengths. As expected, for
small values of $\alpha$ the theoretical expression,
Eq.~(\ref{eq23}), properly accounts for the linear scaling.
Specifically, correlations grow linearly up to approximately
$\alpha \approx 0.25$ [red symbols and line in
Fig.~\ref{fig5}(a)], which represents a frequency adaptation of
about $20\%$ [see red line in Fig.~\ref{fig5}(b)]. This limit is
slightly better than that expected from the upper limit of the
linear scaling in the stationary mean adaptation strength,
$\langle \varepsilon \rangle$ [see Fig.~\ref{fig2}(b)]. For larger
values, higher effects are non-negligible and affect the
correlations in the predicted manner. For a realistic SFA
(adaptation of about $50\%$), the analytical expressions provide a
qualitative agreement regarding the dependence on the parameters
[shape of the curve in Fig.~\ref{fig5}(a) as a function of $\mu$
or $D$], but only a rough estimate of the correct values of the
correlations (cyan symbols and line in Fig.~\ref{fig5}).

\subsection{Spike-count variance reduction}
\indent The presence of negative correlations affects the
spike-count variance. To analyze this effect, it is convenient to
introduce the \textit{Fano factor}, $\text{FF}_{T}$, which relates
the mean and the variance of the spike counts observed in a
temporal window of length $T$, $\langle n_{T} \rangle$ and
$\langle \Delta n_{T}^{2} \rangle$, respectively, as the ratio

\begin{equation}\label{eq27}
   \text{FF}_{T} = \frac{\langle \Delta n_{T}^{2} \rangle}{\langle
   n_{T}\rangle}.
\end{equation}

\indent For point processes, the asymptotic behavior of the Fano
factor reads \cite{cox}

\begin{equation}\label{eq28}
   \text{FF}_{\infty} = \text{lim}_{T\rightarrow \infty} ~ \text{FF}_{T}
   = \text{CV}^{2} ~ \left( 1 + 2\sum_{\text{k}=1}^{\infty} \rho_{\text{k}}
   \right),
\end{equation}

\noindent where $\text{CV}$ is the \textit{coefficient of
variation}, defined as the ratio between the standard deviation
and the mean of the unconditional ISI statistics,

\begin{equation}\label{eq29}
   \text{CV} = \frac{\sqrt{\langle \Delta \tau^{2} \rangle}}{\langle \tau
   \rangle}.
\end{equation}

\indent Combining the preceding equations and given that $\langle
n_{T} \rangle = T/\langle \tau \rangle$, the spike-count variance
reads, in the asymptotic limit,

\begin{equation}\label{eq30}
   \langle \Delta n_{T}^{2} \rangle = \frac{\langle \Delta \tau^{2} \rangle}{\langle \tau
   \rangle^{3}}~\left( 1 + 2\sum_{\text{k}=1}^{\infty} \rho_{\text{k}}
   \right)~T.
\end{equation}

\indent The linear growth in $T$ of the spike-count variance is a
characteristic of a diffusive process (and the reason for the
usefulness of the Fano factor). Inasmuch as the asymptotic limit
is reached, it is useful to analyze the preceding factor, which we
denote $\langle \hat{\Delta n_{T}^{2}} \rangle$,

\begin{equation}\label{eq31}
   \langle \hat{\Delta n_{T}^{2}} \rangle = \frac{\langle \Delta \tau^{2} \rangle}{\langle \tau
   \rangle^{3}}~\left( 1 + 2\sum_{\text{k}=1}^{\infty}
   \rho_{\text{k}}
   \right).
\end{equation}

\indent Equation~(\ref{eq31}) highlights two contributions to the
spike-count variance: a contribution from the FPT statistics
related to a single spiking process, $\langle \Delta \tau^{2}
\rangle / \langle \tau \rangle^{3}$, and a contribution from the
ISI correlations provided by the entire spike train, $\left( 1 +
2\sum_{\text{k}=1}^{\infty} \rho_{\text{k}}\right)$.

\begin{figure}[t]
\begin{center}
\includegraphics[scale=0.35]{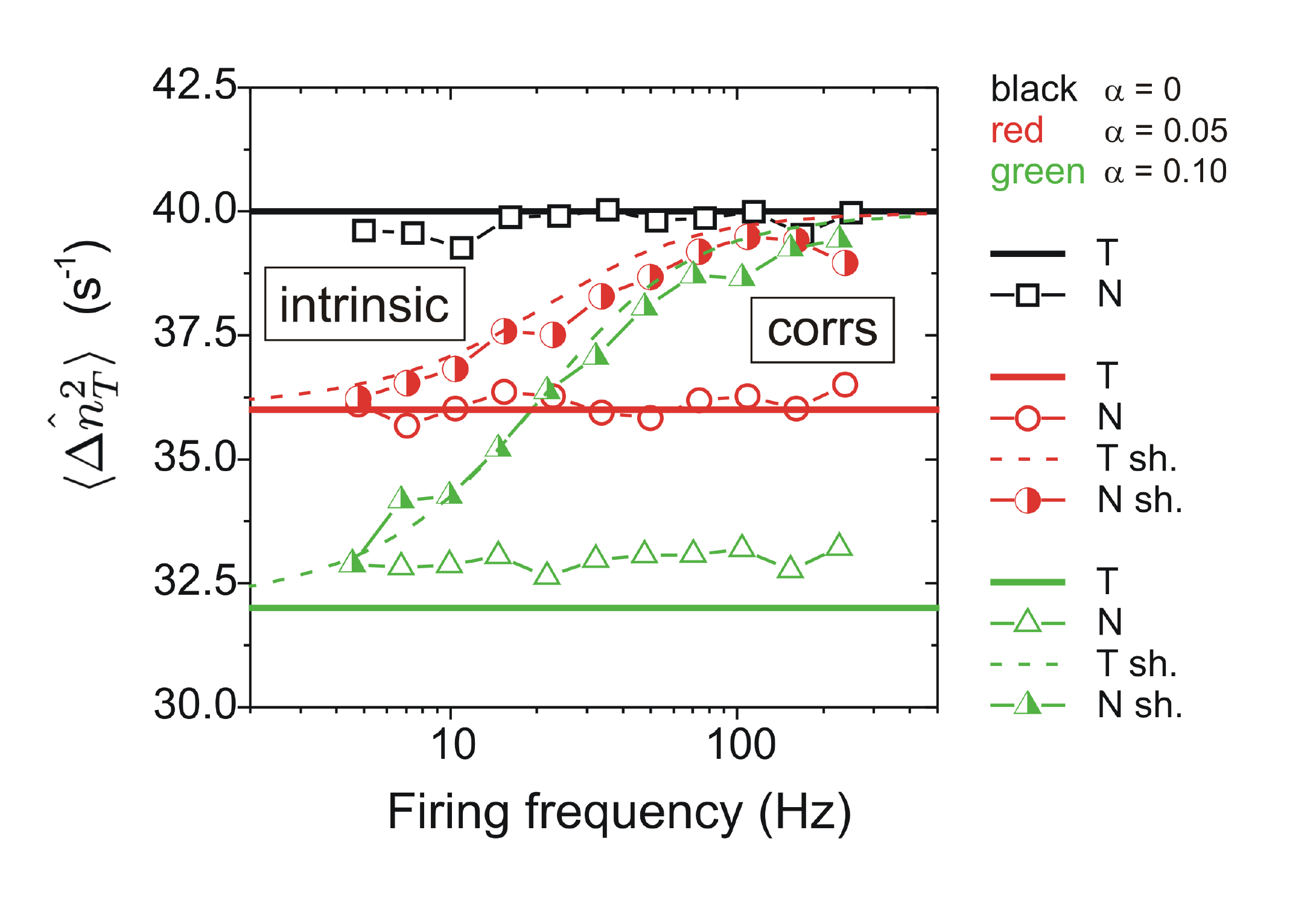}
\caption{\label{fig6} (Color online) Spike-count variability in
the perfect IF neuron model, as a function of the firing
frequency. The neuron without adaptation ($\alpha = 0$) shows a
spike-count variance that does not depend on the firing frequency
[theoretical ($T$) and numerical ($N$) results are represented by
a black thick line and open squares, respectively]. For $\alpha
\neq 0$ ($\alpha = 0.05$ and $\alpha = 0.10$ in red and green,
respectively) there is a reduction in the variance, which also is
independent of the firing rate [theoretical ($T$) and numerical
($N$) results are shown by thick lines and empty symbols,
respectively]. By shuffling the original spike train, from which
the spike counts are drawn, we destroy the correlations. For the
shuffled spike train, theoretical ($T$ sh.) and numerical ($N$
sh.) results are represented by hidden lines and semifilled
symbols, respectively. Parameters: $V_{\text{thr}}-V_{\text{r}} =
1.0$, $\tau_{\text{a}} = 100.0 ~\text{ms}$, $D =
0.02~\text{ms}^{-1}$. The driving parameter $\mu$ was varied from
$0.005$ to $0.2~\text{ms}^{-1}$. Numerical results are presented
as a function of the firing frequency observed from the mean spike
counts. Theoretical results are shown as a function of the firing
frequency computed from $f^{-1} = \langle \tau \rangle_{\phi_{0}}
+ \langle \varepsilon \rangle \langle \tau \rangle_{\phi_{1}}$.
Numerical procedure: to compute the spike-count statistics, the
length of the temporal window $T$ was varied for different values
of $\mu$, checking in all cases that the asymptotic regime was
reached. Numerical results were obtained from a single spike train
with a temporal length $T_{\text{total}} = T~N_{\text{sample}}$,
with $N_{\text{sample}} = 100$. The value of $N_{\text{sample}}$
is low, due to the computational cost of the shuffling procedure.
To improve the accuracy, results are presented as averages over
$N_{\text{repet}} = 500$ repetitions.}
\end{center}
\end{figure}

\indent The effect of the adaptation strength $\alpha$ on the
spike-count variance is twofold; it changes the ISI statistics as
well as the correlations between the ISIs. In the slight
adaptation regime we consider, the contribution to the spike-count
variance due to the correlations is a linear term in $\alpha$.
Explicitly, it is easy to show that

\begin{equation}\label{eq32}
   1 + 2 \sum_{\text{k}=1}^{\infty} \rho_{\text{k}} = 1 + 2
   \frac{\rho_{1}(\alpha)}{1-\phi_{0}^{L}(1/\tau_{\text{a}})},
\end{equation}

\noindent where $\rho_{1}(\alpha)$ is given by Eq.~(\ref{eq23}).
To be consistent with this description, the term provided by the
ISI statistics should be linearized, and reads

\begin{eqnarray}\label{eq33}
   \frac{\langle \Delta \tau^{2} \rangle}{\langle \tau
   \rangle^{3}} = \frac{\langle \Delta \tau^{2} \rangle_{\phi_{0}}}{\langle \tau
   \rangle_{\phi_{0}}^{3}} \hspace{5.1cm}\nonumber\\
   + \langle \varepsilon \rangle ~\frac{\langle \tau^{2} \rangle_{\phi_{1}}
   - 2 \langle \tau \rangle_{\phi_{0}} \langle \tau \rangle_{\phi_{1}}
   - 3 \langle \Delta \tau^{2} \rangle_{\phi_{0}} \langle \tau \rangle_{\phi_{1}}/\langle \tau \rangle_{\phi_{0}} }
   {\langle \tau \rangle_{\phi_{0}}^{3}},
\end{eqnarray}

\noindent where $\langle \varepsilon \rangle$ is given by
Eq.~(\ref{eq11}). Replacing the complete expressions for both
factors [Eqs.~(\ref{eq32}) and (\ref{eq33})] in Eq.~(\ref{eq31}),
and keeping the first order in $\alpha$ [$\langle \hat{\Delta
n_{T}^{2}} \rangle \approx \mathcal{O}(\alpha)$], we obtain

\begin{eqnarray}\label{eq34}
   \langle \hat{\Delta n_{T}^{2}} \rangle = \frac{\langle \Delta \tau^{2} \rangle_{\phi_{0}}}{\langle \tau
   \rangle_{\phi_{0}}^{3}} + \frac{\alpha}{[1-\phi_{0}^{L}(1/\tau_{\text{a}})]\langle \tau
   \rangle_{\phi_{0}}^{3}}\hspace{1.5cm} \nonumber\\
   \times \Bigg[ \langle \tau^{2} \rangle_{\phi_{1}} - 3 \frac{\langle \tau^{2} \rangle_{\phi_{0}} \langle \tau \rangle_{\phi_{1}}}{\langle \tau
   \rangle_{\phi_{0}}}+ \frac{\langle \tau \rangle_{\phi_{1}}}{1-\phi_{0}^{L}(1/\tau_{\text{a}})}\hspace{1.0cm}\nonumber\\
   \times \left( \langle \tau \rangle_{\phi_{0}}
   - 3 \langle \tau \rangle_{\phi_{0}} \phi_{0}^{L}(1/\tau_{\text{a}}) - 2 \frac{d\phi_{0}^{L}(s)}{ds}\rfloor_{1/\tau_{\text{a}}} \right)
   \Bigg].
\end{eqnarray}

\indent It is clear that the independent term on the right-hand
side of Eq.~(\ref{eq34}) corresponds to the asymptotic spike-count
variance (normalized by $T$) of the neuron without adaptation
($\alpha = 0$). Even when general in the regime we consider, the
behavior of this rather complicated expression for $\alpha \neq 0$
is not obvious at all. The expression simplifies enormously for
the case of the perfect IF neuron model. For this neuron, we have

\begin{equation}\label{eq35}
   \langle \hat{\Delta n_{T}^{2}} \rangle =
   \frac{2D}{(V_{\text{thr}}-V_{\text{r}})^{2}}-
   \frac{4D}{(V_{\text{thr}}-V_{\text{r}})^{3}}~ \alpha.
\end{equation}

\indent Surprisingly, this expression for the spike-count variance
reduction does not depend on the driving parameter $\mu$, which
sets the firing frequency. In Fig.~\ref{fig6} we show this
behavior for different values of the adaptation strength. The case
$\alpha = 0$ corresponds to a perfect IF neuron model without
adaptation (black line and squares). As $\alpha$ increases, the
spike-count variance decreases, and this effect does not depend on
the firing frequency set by varying $\mu$. For $\alpha = 0.05$,
theoretical and numerical results agree (red thick lines and empty
circles, respectively); whereas for $\alpha = 0.10$ the flat
reduction is numerically observed, but the appropriate variance
decrease is just approximated by the theoretical prediction (green
thick lines and empty triangles). The same holds true for higher
adaptation strengths (not shown). As expected, since the
expression for $\rho_{1}$ [Eq.~(\ref{eq23})] and the exponential
structure for higher lags [Eq.~(\ref{eq26})] are approximate, the
infinite sum in Eq.~(\ref{eq32}) amplifies a tiny error. In
consequence, even when we have observed that $\rho_{1}$ is
properly described by Eq.~(\ref{eq23}) for the case $\alpha =
0.10$ [Fig.~\ref{fig5}(a)], the expression for the spike-count
variance [Eq.~(\ref{eq35})] is qualitatively good but approximate
[note, however, that there is a significant reduction even for
small values of $\alpha$, and Eq.~(\ref{eq35}) has a moderate
relative error for $\alpha = 0.1$]. The linear spike-count
variance reduction expressed by Eq.~(\ref{eq35}) is unbounded,
which is obviously unreasonable (it enables negative values for
the spike-count variance). Higher-order terms should oppose this
linear reduction, as can be observed in Fig.~\ref{fig6} for the
case $\alpha = 0.10$.\\
\indent Since the spike-count variance reduction for a given
firing frequency arises from the interplay between the FPT
statistics and the presence of correlations, it would be
interesting to disentangle to what extent each effect contributes
to the observed reduction. To analyze this question we shuffled
the spike train, which maintains the first-order statistics (FPT
statistics) destroying ISI correlations. The spike-count variance
reduction for the shuffled spike train is shown in Fig.~\ref{fig6}
as semifilled symbols. The theoretical analog corresponds to
$\langle \hat{\Delta n_{T}^{2}} \rangle$ given exclusively by
Eq.~(\ref{eq33}), since the sum of the SCC values is $0$ [$1 + 2
\sum_{\text{k}=1}^{\infty} \rho_{\text{k}} = 1$ instead of
Eq.~(\ref{eq32})], and it is shown as dashed lines in
Fig.~\ref{fig6}. As expected, theoretical and numerical results
are in good accordance. The behavior observed for each
contribution is reasonable. At low firing frequencies, within each
ISI the exponential evolution of the adaptation process has
decayed, meaning that each $\varepsilon_{\text{n}} = \alpha$ and
correlations disappear. Correspondingly, the spike-count variance
reduction is given exclusively by the FPT statistics (this case
corresponds to $\tau_{\text{d}} \rightarrow 0$ in
\cite{urdapilleta, lindner2004b}), and we denote this regime as an
\textit{intrinsic} reduction in Fig.~\ref{fig6}. At large firing
frequencies, the adaptation process within a single ISI is
essentially constant (i.e., $\tau_{\text{a}} \rightarrow \infty$).
In this case, the collection of ISIs satisfies a quasistatic
approximation \cite{urdapilleta2009}, and the spike-count variance
of the shuffled spike train is indistinguishable from the case of
no adaptation. However, in this limit, correlations decay very
slowly [$\phi_{0}^{L}(1/\tau_{\text{a}}) \rightarrow 1$; see
Fig.~\ref{fig3}(c)], and even when each SCC is small because
$\rho_{1}$ is, the sum of correlations accumulates over many lags,
building up a finite nonvanishing value. In Fig.~\ref{fig6}, we
denote this range as a reduction due to correlations
(``corrs'' in the figure).\\
\indent Equation~(\ref{eq35}) shows that the spike-count variance
reduction is independent of the firing frequency, for the perfect
IF neuron model. As shown in the previous paragraph, in the limit
of low as well as large firing frequencies, the mechanisms that
account for the reduction are different. The fact that both
effects influence the spike-count variance to the same extent, and
furthermore, that these mechanisms exactly compensate each other
at intermediate firing frequencies is intriguing. Obviously, these
results hold for this particular IF model. It would be interesting
to analyze the contributions in other IF models; we expect
analogous spike-count variance reductions and the same limit
behaviors, but not an independence on firing frequencies (in
general, in other models the spike-count variance for $\alpha = 0$
depends on the firing frequency).

\section{Discussion and concluding remarks}
\indent In this work we have analyzed the onset of correlations
for a general neuron model, where an external input as well as an
internal spike-based adaptation current drive the membrane
potential. The external current is composed by a static input and
fast fluctuations. For this system, the dependence of the
adaptation current on the past history, through the initial states
of the adaptation process, facilitates the development of
correlations between successive ISIs \cite{prescott2008,
benda2010, wang1998, liu2001, avila2011}. In the regime of slight
adaptation, we have shown that correlations share a general
structure across different models. By means of a hidden Markov
model, we have explicitly derived the dependence of the SCC on
different properties of the FPT statistics corresponding to the
underlying time-inhomogeneous stochastic process,
Eqs.~(\ref{eq23}) and (\ref{eq26}). In this regime, for
one-dimensional models such as IF neuron models, the necessary
properties are given \cite{urdapilleta, comment1, lindner2004b,
lindner2005b}. For any other (high-dimensional) model, whenever a
slight exponential time-dependent current smoothly reshapes the
FPT statistics in comparison to the unperturbed case (analogously
to Fig.~1 in \cite{urdapilleta}), the expressions derived here
apply. The geometric structure and exponential decay for the SCC
is surprisingly simple and general for the scenario considered
here. This kind of structure was first observed by Lindner and
Schwalger for successive escapes of an overdamped Brownian
particle in a randomly modulated asymmetric double well, which can
be modeled as transitions between discrete states
\cite{lindner2007, schwalger2008b}. Posteriorly, these authors
extended their results to a situation with multiple internal
states \cite{schwalger2010b} and, in particular, studied the case
of negative correlations in neurons with inhibitory feedback
(resembling current-based adaptation), with an appropriate scheme
of transitions. Even when general and very promising, a
quantitative evaluation of correlations in adapting neurons under
this framework requires a procedure for the estimation of
transition rates, in a discrete version of adaptation, obtained
from dynamical models and/or experimental data. The results
obtained here are in qualitative agreement to those obtained by
Schwalger and Lindner \cite{schwalger2010b}, which implies that
this estimation procedure could be an interesting topic to
study.\\
\indent The development of negative correlations in successive
ISIs in a spike train influences the encoding capabilities of a
neural system \cite{avila2011, nawrot2007, nawrot2010,
farkhooi2009, farkhooi2011}. Here, we have analyzed the decrease
in the asymptotic spike-count variance due to the intrinsic
variability reduction of the FPT statistics and the presence of
correlations [Eqs.~(\ref{eq31})-(\ref{eq34})] in comparison to the
case without adaptation. In particular, for the perfect IF neuron
model the decrease is extremely simple and does not depend on the
firing frequency, Eq.~(\ref{eq35}). This analysis is theoretically
important, but it should be put in the proper context. The
spike-count variance reduction given by Eq.~(\ref{eq31}) is valid
for the asymptotic limit, which can be unfeasible in real neurons,
especially at low firing rates. For systems operating with finite
temporal windows, the framework presented here should be extended
by using the complete formalism derived by van Vreeswijk in
\cite{vreeswijk}, and finely used by Farkhooi \textit{et al.} to
analyze a population scheme \cite{farkhooi2011}. In this case, the
spike-count variance will be a function of the length of the
temporal window used to compute the statistics, and obviously, the
results presented here should agree in the limit of large windows.
This behavior was outlined by Chacron \textit{et al.} in a
different version of adaptation (see results of the model without
slow noise in Fig.~4 of \cite{chacron2001}). That work also
highlighted a possible explanation to the interesting finding made
by Ratnam and Nelson \cite{ratnam2000}, where the spike-count
variance exhibits a minimum as a function of the length of the
counting window (a possible behaviorally important phenomenon). In
their work, Chacron \textit{et al.} demonstrated that a slow
external noise leaves relatively intact the short-range
correlations, while destroying negative correlations at large
lags, giving rise to small and slowly decaying positive
correlations which dominate the asymptotic regime. On the other
hand, across different neural systems it has been observed that
the only significant SCC corresponds to the lag 1 \cite{avila2011,
farkhooi2009}, which reinforces the idea that a slow external
noise would be an important part of the incoming signal, in
addition to fast fluctuations. Theoretically, the analysis of the
spike-count variance for IF neuron models driven by slow
fluctuations were successfully carried out via a quasistatic
approximation \cite{middleton2003, schwalger2008a}. This suggests
that the hidden Markov model used here to model negative
correlations could be extended with a similar quasistatic
approximation in order to include slow fluctuations in the
analysis.

\section{ACKNOWLEDGMENTS}
\indent The author thanks In\'es Samengo for a critical reading of
the manuscript. This work was supported by the Consejo de
Investigaciones Cient\'ificas y T\'ecnicas de la Rep\'ublica
Argentina.


\begin{thebibliography}{99}
\bibitem{comment1}
In \cite{urdapilleta} we have found the FPT density function for
the perfect IF model (Wiener process) with the exponential
time-inhomogeneous drift supporting the subthreshold dynamics of
the adaptation current. As pointed out in the conclusions of
\cite{urdapilleta}, the system of recurrence equations can be
extended to other diffusion processes [here characterized by $g(V)
\neq 0$]. In this case, the homogeneous part of the equations to
solve, Eqs.~(8) and (9) in \cite{urdapilleta}, changes
respectively. Given that the homogeneous part is the same for all
order terms, the formal solution of each term is obtained from the
Green's function associated to the unperturbed system and the
corresponding source.
\end{thebibliography}
\end{document}